# BUSINESS PROCESS MINING APPROACHES: A RELATIVE COMPARISON

## Saiqa Aleem[1], Luiz Fernando Capretz[2], Faheem Ahmed[3]

[1,2]Department of Electrical & Computer Engineering, Western University, London, ON (Canada)

[3]Department of Computing science, Thompson River University, Kamloops, BC (Canada)

## ABSTRACT

*Recently, information systems like ERP, CRM and WFM record different business events or activities in a log named as event log. Process mining aims at extracting information from event logs to capture business process as it is being executed. Process mining is an important learning task based on captured processes. In order to be competent organizations in the business world; they have to adjust their business process along with the changing environment. Sometimes a change in the business process implies a change into the whole system. Process mining allows for the automated discovery of process models from event logs. Process mining techniques has the ability to support automatically business process (re)design. Typically, these techniques discover a concrete workflow model and all possible processes registered in a given events log. In this paper, detailed comparison among process mining methods used in the business process mining and differences in their approaches have been provided.*

## I. INTRODUCTION

With the emerging globalization and competent environment, enterprises are required to improve their existing business processes. For that purpose, combination of data mining and machine learning techniques are introduced to workflow field, i.e., process mining [1]. Due to fast paced competitive market, it become essential that enterprise have continues and insightful feedback on how business process actually being executed within the enterprise. In short, enterprise requires a good analysis tool to get insight into its business processes. The requirement of an enterprise to know how actually processes happen in real world is a major driver behind the development and increasing use of process mining approaches.

The goal of business process mining or in short process mining is to automatic construction of process models from event log. The extracted model (e.g. in form of Petri-Net) is the explanation of observed behavior in the event logs. For mining purpose, event logs are used as a starting point. In some systems events log referred to as history, audit trails and transactional logs. It is possible to record events in such a way that each event may refer to a task or a case. The processes or cases in an event log can be looked from different perspectives [2]: (a) process perspective (How?) focuses on the control flow for example ordering of activities [3], (b) organization perspective (Who?) focuses on the performers field, i.e. how they are related and which performers are involved; and (c) the case perspective (What?) focuses on case properties to establish relation between them, e.g. characterization of cases based on path in the process, performers performance or their utilization.

The rest of the paper is organized as follows: Section 2 briefly describes about the different approaches utilized in process mining, Section 3 discusses and compares those approaches in detail and finally section 4 concludes the paper.

## II. BUSINESS PROCESS MINING APPROACHES

Business process mining exploits the information recorded in events logs and portrays a family of a-posteriori analysis approach. To analyze processes in event logs process mining is a suitable technique. The proposed approaches for process mining can be distinguished in to three types [4]: discovery, conformance and extension as shown in the above Fig. 1.

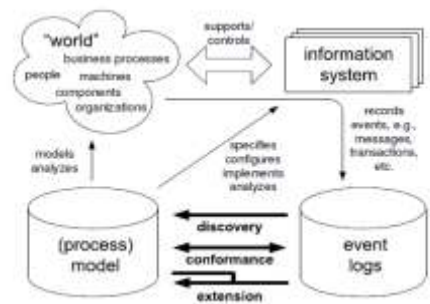

**Fig: 1Three Types of Process Mining: (1) Discovery, (2) Conformance, And (3) Extension [4]**

Discovery based process mining techniques extract information related to data from event log only.

Conformance based techniques verify that if actual processes in enterprise follow prescribed behaviors or rules.

Extension base techniques takes existing model as input and enhance that model based on the information extracted from event logs. There is also a-priori model.

Process mining can be described as a sub-domain of data mining. In process mining, event logs can consist of information about the attribute of cases and actual flow or route of the tasks by case. Traditional data mining approaches mine the decision rules that predicts the flow or route of a case where as process mining focuses on the mining of process model. In section III, we describe the different approaches used for business process mining.

## III. APPROACHES

The research in process mining dedicated by many scholars and so far they have made lots of achievements and improvements in the field of process mining. Earliest work related to process mining investigated by Cook and wolf in the context of software engineering processes [5].

Their proposed work was not directly related to business processes discovery but they introduced the basic thoughts about process mining. Their proposed algorithm deals with noise and parallel structure [1]. Later they also worked on conformance of models that is based on concurrent processes of probability [6]. The idea of process mining in the context of workflow management was first introduced in [7]. He is the one who named the technology officially as process mining and proposed an algorithm. They stated that algorithm that computes conformal graph with complexity $O(m^{x}n^3)$. It deals with noise and parallel structure.

A meta model for process mining proposed by Dongen and Van der Aalst [8]. A new format based on XML called MXML. Their work is more programmatic and driven by concrete tools. They claimed that this framework will help researchers in implementing new process mining techniques without caring about the information systems, the event logs generated by. For process mining, broad variety of approaches/algorithms does exists and for better understanding they are further classified as follows.

### 3.1 Deterministic Mining Approach

The approaches that belong to this category only produce reproducible and defined results. The α-algorithm based approaches can be classified under this category. Van der Aalst *et al.* [9] proposed α-algorithm that can mine any workflow and is represented in SWF-net i.e. sub class of Petri nets. To explain SWF-net is out of the scope of this paper. If further information is required on SWF-net refer to [9]. It was one of the first approaches to take concurrency into account (i.e., explicit causal dependencies and parallel tasks). Algorithm first analyzes the event log then establishes various dependency relations between tasks. Relations between tasks considered as casual and also describe the sequence of the tasks. Fig. 2 showed that these types of constructs cannot be mined.

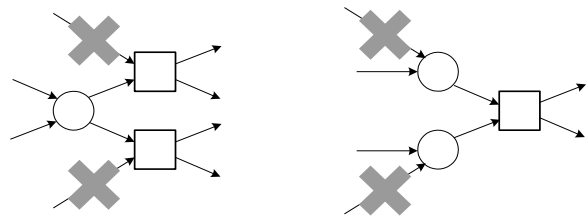

**Fig. 2 Construct Cannot Be Mined In A-Algorithm [9]**

### 3.2 Heuristic Mining Approaches

Heuristic techniques used by process mining algorithms to better deal with noise and incompleteness [9]. It involves three mining steps, first of all it constructs the dependency/frequency table (D/F-table), secondly mine the basic relations out of the D/F-table (R-table) and then finally reconstruct the WF-net out of the R-table.

van der Aalst [10] demonstrated the applicability of process mining in real-world. They used this heuristic approach to deal with problems in process mining. They reported that most of the existing systems do not focus on casual and dynamic dependencies in process and organizations. To handle that problem they used heuristic approach [9]. In order to capture organization perspective, they utilized Social Network Analysis (SNA) [11] techniques and embeds the MiSoN [2] functionality in the framework for analyzing relationships. The model is represented in terms of Petri-Nets. The process mining refers to logical structure of the process model and performance issues such as flow time. They used ProM framework (Plug-able environment) to cover the all three perspective in an organization. Five different types of plug-ins are allowed with in this framework: mining plug-in, export plug-in, import plug-in, analysis plug-in and conversion plug-in. ProM uses MXML format for process mining. To achieve a richer understanding of the processes in an organization, it is worthwhile to combine different mining perspectives as showed in this case study. Number of limitations of process mining also showed in this case study. For example, only those events were considered which are logged, privacy issues are also there.

### 3.3 Inductive Approach

The purpose of this approach is to acquire the business process models and their adaptation to the changing requirements by identifying the best HMMs (Hidden Markov Models) that showed the process model. Inductive approach involves two steps (i) Induction step and (ii) Transformation Step. Induction step involves creation of Stochastic Task Graph (SAG) from event log. The transformation step is responsible for the synchronization of structures of event log instances, generation of synchronized structures and generation of process model. Herbst and Karagiannis [12] dealt with the duplicate tasks and their proposed algorithm based on inductive approach two steps. First step is induction step same as described above and in second step SAG is generated. That graph is then transformed into blocked structured model in ADONIS definition language (ADL). The developed tool is called InWoLvE that takes many parameters but it requires proper parameters to improve mining efficiency and quality. Schimm [13] proposed an approach that deals with the hierarchal structural workflow model and used inductive bias. That model deals with the splits and joins and it extract accurate model from event log. He demonstrated that method by using detailed example and also developed a tool process miner. Wainer *et al*. [14] proposed approach based on process algebra and is used to derive final model, they utilized the existing models. It is type of Extension in process mining. The main idea is to rewrite the existing models with new instances, it is complex in implementation and have high time and space complexity.

### 3.4 Genetic Mining Approach

Genetic algorithm [15] can be used to mine process models out of events log. First of all, it creates the initial population of workflows. For that purpose, it builds the causal Metrics that contains the relations among tasks. In second step it calculates the Fitness of each individual. The main idea of using is to benefit the individuals that can parse more frequent material in the log. To calculate the fitness, a continuous semantics parser and register problems technique utilized.

$$F(L,CM) = \frac{allParsedActivities(L,CM) - punishment}{numActivitiesLog(L)}, \text{ where}$$
$$punishment = \frac{allMissingTokens(L,CM)}{numTracesLog(L) - numTracesMissingTokens(L,CM)+1} + \frac{allExtraTokensLeftBehind(L,CM)}{numTracesLog(L) - numTracesExtraTokensLeftBehind(L,CM)+1}$$

The above function F is used to find out that how an individual is fit to a log. In third step of the algorithm, for the creation of next generation genetic operators such as Cross over and Mutation were used. Cross Over operator recombine the existing population, cross point are the task, calculate the probability of cross over and then subsets can be swapped. In Mutation operator, it introduces new material in the population then every task of an individual can be mutated based on mutation probability. Medeiros *et al*. [15] used this algorithm as plug-in in PROM framework and performed experiments on the simulated data. The results showed that this algorithm found all possible business process models that could parse all the trace in the event log.

### 3.5 Clustering Based Mining Approaches

Process mining approaches based on clustering have been developed in order to solve the problem of complex process model like "spaghetti shape". Clustering based approaches generate similar process models to explain the behavior of single process and cluster them together as compared to traditional approaches generate only one single mode [16]. Some of the approaches are based on Bag of Activities and K-gram model, usually transom each process instance into a vector in order to analyze each process. These methods do not have information about the order of performed activities and their context. Song *et al*. [17] have proposed that combination of different perspective can be defined by vector such as data, control flow and performance etc. and tried to get better results than previous approaches. The proposed approach is not able to solve the issue of lack of context. Another approach, Edit_Distance, tried to solve the issue by comparing two process instances and assigned the cost by calculating difference between two sequences [16]. There is another method called trace clustering [18] that utilizes robust set of features for calculating process instances similarity in order to create different clusters. Bose *et al*. [19] proposed method presented the general schema of features and statistical technique that detect the changing points and identify changed regions from control flow perspective. Luengo and Sepulveda [20] extend the work of Bose *et al*. [19] by adding time feature and resultant clusters shared the structural similarity as well as temporal proximity. Finally, Ali *et al*. [21] extended business intelligence to the healthcare domain.

## IV. DISCUSSION

In context of process mining many efforts has been made but still most of the challenges described by van der Aalst [22] were not addressed. Current approaches are facing difficulties when mining business process that contain non-trivial constructs or when log contains noise. Most of the approaches assume that the event logs are complete and noise free but this assumption is unrealistic. The list of issues that can be a challenging problem in order to construct a process model can be:

**4.1 Mining Hidden Tasks**

The log may contain some hidden task or invisible task. These tasks can be used only for routing purpose and are not recorded in the log. Fig. 3 shows that if a model contain hidden task then the resultant model is not the accurate representation of process model. The approach proposed by Song *et al*. [23] is the only one that handles the problem of hidden tasks.

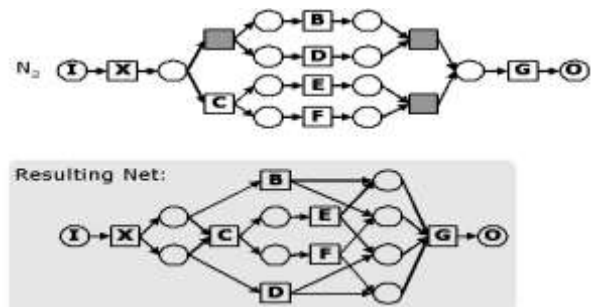

**Fig: 3 An Example of Hidden tasks Translation [24]**

**4.2 Mining Duplicate Tasks**

Duplication of one task may occur in event log as depicted in Fig. 4. If multiple transition have same label then they appear as duplicate task and they are treated as single task by process mining approach. Duplicate task tackle by α-algorithm [9] do not consider the redesign problem. Duplicate Tasks are the one issue that is addressed by other researchers [12, 13, 25, 23].

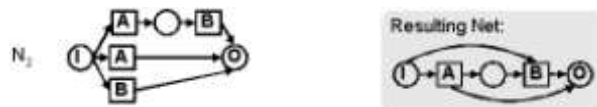

**Fig: 4 An Example of Duplicate Task Translation [24]**

**4.3 Mining loops**

Event logs may contain different types of loop. Those loops can be of length one or two they can be sometimes called as short loop. When dependency relationships are extracted from workflow, for the length one loop as shown in Fig. 5. B>B and not B>B implies B→B it is impossible to detect. For length two loop as shown in Figure 5 A>B and B>A implies A∥B and B∥A instead of A→B and B→A. These types of construct cannot handle by α-algorithm [9]. Other approaches [13, 14] are also covered this problem in their algorithms.

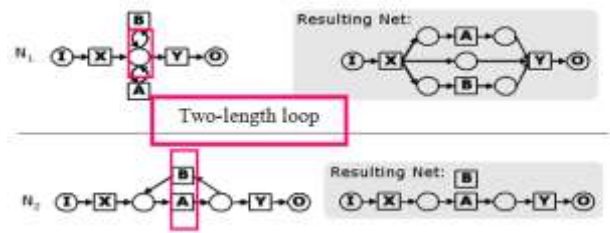

**Fig. 5 An Example of Mining Loop Translation [24]**

### 4.4 Mining Non-Free-Choice Constructs

If one task is the input for other two different tasks then that transformation results into non-free choice construct. These types of construct combine choice and synchronization. This type of constructs are difficult to mine by mining algorithm as the choice is non-local and algorithm has to remember earlier events. This problem is addressed by very few researchers. Huang and Zhang [26] and Song *et al*. [23] solve this issue by proposing γ+ algorithm. Other than above mentioned challenges there are some other construct problems, noise and completeness issues that are also not easy to handle in process mining algorithms. Noise can appear if the tasks somehow incorrectly logged or event log reflects exceptional cases. The α – algorithm has so many limitations as discussed before, to solve those problems many researchers extend that algorithm.

Algorithms like heuristic based, α-algorithm based or inductive approach proved to be more significantly time efficient and linear in the size of the log. However, on real-life data such algorithms do not perform well. So, to deal with real life applications more advanced algorithms needed like genetic approach.

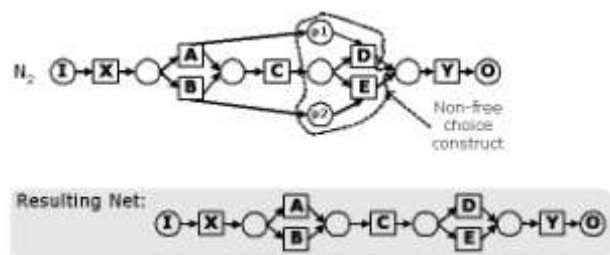

**Fig. An example of Non- Free choice construct Translation [24]**

But main disadvantage of genetic approach is its time and space complexity. To overcome high computational problem, Bratosin *et al*. [27] proposed distributed genetic approach. They used the coarse grained distributed variant of the genetic miner i.e. plug-in included in Prom Tool and their results showed that it improved time complexity somehow.

From the above discussion, it is clearly seems that none of the technique addresses all of the issues. Non free choice constructs and consideration of noise in the event log is considered by few of the researches. Issues related to duplicate tasks, hidden tasks and non-free choice constructs utmost while others only handle duplicate tasks or loops. Most of the approaches do not handle noise issue. Their assumption of noise free and completeness of event logs is unrealistic. So, we need approach or combination of approaches to address all of those issues in order to get efficient and qualitative process mining.

## V. CONCLUSION

In an organization due to many reasons actual work can deviate from process definitions. Therefore, it becomes irresistible for organizations to discover these deviations in order to improve their processes. Process mining allows identification of processes from event logs and the discovery of differences between the prescriptive process model and the real world process executions. This paper presented the detail comparison of approaches used for process mining. Comparison provided between them also covered the aspect that either they address technical issues or not. To get efficient and qualitative results of business process mining, researchers have to look for comprehensive standard technique or combination of techniques that is based on realistic assumptions. Process mining is a stimulating topic both from a practical and scientific perspective, and it should be further exploited through research in software analytics.

## VI. ACKNOWLEDGMENT

The author would like to thank Dr. Charles Ling (Computer Science Department, Western University) for his constructive comments which contributed to the improvement of this article as part of his course work.